\documentclass{article}
\usepackage[utf8]{inputenc}
\usepackage{xcolor}
\usepackage[sorting=none]{biblatex}%[sorting=none]
\usepackage{graphicx}
\usepackage{float}
\usepackage[english]{babel}
\usepackage{tablefootnote,footnote,csquotes}
\usepackage{pdfpages} 
\usepackage{array, caption, tabularx, booktabs,authblk,pdflscape,tabularray,geometry}%ragged201e
\usepackage[strict]{changepage}
\usepackage{helvet}
\usepackage[symbol]{footmisc}

\title{Applying the causal roadmap to longitudinal national Danish registry data: a case study of second-line diabetes medication and dementia}
\date{}
\author[1]{Nance, N.\footnote{these authors contributed equally}}
\author[1]{Mertens, A.*}
\author[2]{Gerds, T.}
\author[1]{Wang, Z.}
\author[2]{Torp-Pedersen, C.}
\author[1]{van der Laan, M.}
\author[3]{Kvist, K.}
\author[2]{Lange, T.}
\author[2]{Zareini,B.}
\author[1]{Petersen, M.}
\affil[1]{University of California, Berkeley}
\affil[2]{University of Copenhagen}
\affil[3]{Novo Nordisk}

\begin{document}

\maketitle

\begin{abstract}
%Here we illustrate the causal roadmap, a formal framework for causal and statistical inference, through an applied longitudinal analysis of the effects of second-line diabetes drugs on dementia risk using Danish National Registry data. Specifically, we compared the five-year cumulative risk of dementia between individuals continuously using glucagon-like peptide 1 receptor agonists (GLP-1RAs), a newer class of  drugs, versus other second-line drug. Weestimated the target statistical parameters with longitudinal targeted maximum likelihood estimation to minimize estimation assumptions and control for time-dependent confounding. We specify the analysis using outcome-blind simulations. We found a modest protective effect of GLP-1RAs compared to other second-line treatments, and attenuated effects compared to specific active comparators. This case study outlines the first application of the longitudinal causal roadmap to Danish registry data.
%maya edits below:
The causal roadmap is a formal framework for causal and statistical inference that supports clear specification of the causal question, interpretable and transparent statement of required causal assumptions, robust inference, and optimal precision. The roadmap is thus particularly well-suited to evaluating longitudinal causal effects using large scale registries; however, application of the roadmap to registry data also introduces particular challenges. In this paper we provide a detailed case study of the longitudinal causal roadmap applied to the Danish National Registry to evaluate the comparative effectiveness of second-line diabetes drugs on dementia risk. Specifically, we evaluate the difference in counterfactual five-year cumulative risk of dementia if a target population of adults with type 2 diabetes had initiated and remained on GLP-1 receptor agonists (a second-line diabetes drug) compared to a range of active comparator protocols. Time-dependent confounding is accounted for through use of the iterated conditional expectation representation of the longitudinal g-formula as a statistical estimand. Statistical estimation uses longitudinal targeted maximum likelihood, incorporating machine learning. We provide practical guidance on the implementation of the roadmap using registry data, and highlight how rare exposures and outcomes over long-term follow up can raise challenges for flexible and robust estimators, even in the context of the large sample sizes provided by the registry. We demonstrate how simulations can be used to help address these challenges by supporting careful estimator pre-specification. We find a protective effect of GLP-1RAs compared to some but not all other second-line treatments. 

\end{abstract}

\clearpage
\section{Introduction}\label{sec1}

There has been a significant increase in the amount of existing observational or ``real world’’ health data both collected by and aggregated for researchers over the past decades. This type of data has become increasingly important in medical decision-making, as it allows researchers to answer questions that would not be feasible to address in a randomized trial due to financial constraints, ethical considerations, or logistical challenges. Electronic health record databases, with their large sample sizes and long follow-up times, enable the study of rare outcomes, and realistic treatment usage. However, because observational data are particularly susceptible to confounding, real-world evidence is often used only for associational analyses and causal claims are discouraged. \cite{hernan_c-word_2018,parra_consistency_2021,amastyleinsider_use_2017} Further, while measured confounders can be adjusted for, standard statistical estimation approaches that rely on pre-specified parametric models may fail to fully adjust for measured confounders (introducing bias that does not decrease with greater sample size), while \emph{ad hoc} model adaptation based on data exploration undermines the basis of robust statistical inference.\cite{laan_targeted_2011}

The causal roadmap provides a structured framework for navigating these challenges and generating robust evidence for causal questions, when appropriate assumptiosn are met.\cite{petersen_causal_2014}  Specifically, the causal roadmap provides guidance for the following steps: 1) translation of a causal question, such as the comparative effectiveness of two longitudinal treatment protocols, into a formally defined \emph{causal estimand}; 2) explicit statement of the \emph{observed data} and knowledge about the processes that gave rise to it (\emph{causal model})  3) \emph{identification}, or the formal translation of the causal estimand (which describes an ideal hypothetical experiment) into a \emph{statistical estimand} (a function of the observed data distribution that can be estimated) under explicit causal assumptions;  4) pre-specification of a \emph{statistical model} that avoids any unsubstantiated statistical assumptions, and a \emph{statistical estimator} selected to provide the best expected performance (e.g., nominal 95\% CI coverage, minimum variance); and 5) support for \emph{interpretation} in the context of these assumptions, including sensitivity analyses.

Large, comprehensive, and representative longitudinal datasets like national registries offer advantages for evaluating causal effects; they allow for generalizable study populations and investigation of the real-world effectiveness of long-term treatments on rare outcomes. Their use also introduces challenges, making implementation of the casual roadmap simultaneously more complex and arguably even more essential to ensure robust inferences. 

The first challenge arises because the full longitudinal counterfactual interventions of interest must be specified, including not only initial treatment assignments, but also any post-baseline restrictions on treatment modification or drop-in, as well as any hypothetical interventions on treatment compliance, censoring or measurement.\cite{petersen_applying_2014} Specification of a ``target trial’’\cite{hernan_using_2016}  is one commonly used device to support specification of a causal estimand. Of note, however, specification of post-baseline ``interventions’’ or protocols, including those to enforce compliance or to prevent censoring, need not correspond to any realistic intervention that might be accomplished in an actual trial.

Secondly, both the observed data and confounding structures in longitudinal registry data are complex, and thus potential identification of the target causal estimand in such studies requires a more complex statistical estimand, such as the longitudinal g-formula.\cite{laan_targeted_2018} This, in turn, necessitates more complex statistical estimation procedures capable of adjusting for time-varying confounders. Finally, many of the advantages of using registry data---including the ability to study the effects of prolonged treatments on rare outcomes and to capture of a wide range of potential and time-dependent confounders---also introduce additional challenges to estimation. Flexible machine learning approaches to covariate adjustment are paramount to making optimal use of complex longitudinal data captured in a registry, but must be implemented such that statistical inference is preserved.\cite{laan_targeted_2018} Further, even with initially large sample sizes and common treatments, the number of patients who comply with long-term protocols can dwindle quickly, resulting in practical positivity violations; examples of this include longitidinal inverse probability of treatment weighting (IPTW), parametric g computation, and longitudinal maximum likelihood estimation.\cite{petersen_diagnosing_2012} Finally, the unique ability of registries to evaluate effects on rare outcomes can introduce new challenges to estimator performance. Careful estimator specification and benchmarking can address many of these challenges; however, in order to provide a firm basis for statistical inference, the full estimation procedure must be pre-specified.

Motivated by both the utility of the causal roadmap for longitudinal registry analyses and the practical challenges that arise in its application, in this paper we present a detailed case study. Through it, we provide practical guidance on implementation of the longitidinal causal roadmap, with a particular emphasis on the use of simulations to guide pre-specification of a statistical estimator that integrates machine learning. Specifically, we utilize the roadmap to evaluate the long-term cumulative causal effect of second-line diabetes medications on dementia risk among a cohort of diabetes patients using the Danish National Registry. The Danish National Registry is one of the longest-standing national registry databases, and has expansive longitudinal data dating back several decades.\cite{schmidt_danish_2015} Due to its large size, rich comorbidity and concurrent medication measurements, and extended follow-up time, the Danish registry is well-suited for examining rare and longer-term exposures and outcomes. It also has a high degree of diagnostic reliability for some but not all diseases, with a positive predictive value of 0.96 and 0.98 for diabetes and dementia, respectively.\cite{thygesen_predictive_2011}  Dementia affects around three percent of the older Danish population, but several studies suggest that the true prevalence is higher.\cite{noauthor_dementia_nodate-1} Dementia is particularly burdensome to the diabetic population.\cite{biessels_risk_2006} Continually emerging evidence suggests newer second-line diabetes medications may hold promise in preventing neurodegeneration(eg. \cite{ballard_liraglutide_2020,vadini_liraglutide_2020,norgaard_treatment_2022}); however, some newer medications are less historically common relative to more established therapies.\cite{knudsen_changes_2020} Due to the relatively uncommon nature of exposure and outcome of interest and the need for prolonged follow up, examining this relationship without the use of a large data source like the Danish registry would be infeasible.

With this motivation, we evaluate the difference in the counterfactual five-year cumulative risk of dementia if a target population of adults with type 2 diabetes had initiated and remained on GLP-1 receptor agonists (GLP-1RAs), a relatively new class of drugs, compared to a range of active comparator protocols. We utilize a non-parametric statistical model that avoids unsubstantiated assumptions, and is thus known to contain the true observed data distribution; this contrasts with commonly employed approaches to statistical estimation in which a more restrictive statistical model (such as a main term logistic model on the propensity score) is adopted from convenience or habit. The use of a large semi- or non-parametric statistical model, as dictated by the Roadmap, provides crucial protection against bias due to both model misspecification and post-hoc data exploration. It does, however, set a high bar for choice of a statistical estimator. Specifically, the pre-specified statistical estimator must be able to provide precise and robust estimates of the treatment effect without reliance on simplified lower-dimensional parametric models. 

Time-dependent confounding is accounted for through use of the iterated conditional expectation expression of the longitudinal g-formula as the statistical estimand.\cite{bang_doubly_2005,wen_parametric_2021} Longitudinal targeted maximum likelihood estimation (L-TMLE)\cite{petersen_applying_2014,laan_targeted_2018} provides a general estimation approach for these causally derived estimands, while integrating ``double’’ machine learning (in both a series of outcome regressions and in estimates of propensity score), in order to flexibly incorporate covariate adjustment.\cite{laan_targeted_2011}  However, full pre-specification of a TMLE for this estimand requires several key decisions. These decisions can have profound impacts on finite sample estimator performance, particularly in the challenging setting of large scale registry analysis of  long-term treatment effects on rare outcomes. Further, given the challenges of rare outcomes in our case study, an investigator may also wish to verify whether the use of a TMLE is warranted, or whether a simpler estimator such as in inverse probability of treatment (IPTW) estimator, despite having theoretical shortcomings, might provide to be an equivalent or superior finite sample estimate.

Our case study provides practical guidance on the implementation of the roadmap in large-scale registry data. In particular, we highlight how rare outcomes can raise challenges for flexible and robust estimators such as L-TMLE, even in the context of the large sample sizes provided by the registry, and discuss estimation approaches to addressing these challenges, including careful specification of the machine learning library and selection of an appropriate variance estimator. We show how simulations can be used to help pre-specify the estimation algorithm (a prerequisite for valid statistical inference) in order to optimize the expected performance of the statistical estimator (in terms of confidence interval coverage and precision). As we discuss, simulations can also be useful when working with registry data that can be difficult to access for logistical reasons, prompting a further need for simulated data. There are examples of previous analyses of Nordic registry data using a target trial framework (eg. \cite{olarte_parra_trial_2022,rossides_infection_2021,gavoille_investigating_2023}).  However,
%these analyses do not follow a causal roadmap, which can help guide analysis specification. For example, one study reports the hazard ratio, which is generally difficult to interpret causally.\cite{hernan_hazards_2010} To t
to the best of our knowledge, this is among the first studies to apply the causal roadmap to a national registry of historic data on over 200,000 diabetes patients.\cite{international_diabetes_federation_denmark_nodate}

The paper is organized as follows. In Section 2, we step through the application of the causal roadmap to our case study, evaluating the long-term effect of GLP-1RA use on incident dementia risk using the Danish National Registry, up to the step of statistical estimation. We  illustrate opportunities, challenges, and practical responses at each step of the roadmap. In Section 3, we specify the statistical model and detail the estimation procedure, including the simulation study, which was used to pre-specify the analysis. In particular, we provide a detailed overview of key estimation choices in the implementation of our longitudinal targeted maximum likelihood estimator, and the use of simulations to guide these choices. In Section 4, we present the results and interpretation of our analysis, including sensitivity analyses. In Section 5, we discuss the findings and the implications of this work.

\section{Case study of the longitudinal causal roadmap: evaluating the long-term effect of GLP-1RA use on incident dementia risk using the Danish National Registry}
\subsection{Background and motivation}

Type two diabetes mellitus (T2DM) is a cardio-metabolic disease that affected 462 million people globally in 2017 and is increasing in prevalence each year.\cite{khan_epidemiology_2020} It is characterized by elevated blood sugar levels, which over time can strain not only the cardiovascular but also the nervous system.\cite{cdc_what_2020} Dementia is a progressive neurodegenerative disease characterized by impaired cognitive functioning that interferes with daily life. T2DM has been shown to accelerate both brain aging and neurodegeneration, which are antecedents to dementia.\cite{biessels_risk_2006,umegaki_type_2014} While the exact mechanisms are an active area of research, hypothesized mechanisms include a complex mixture or pathways---including defective insulin signaling, metabolic/mitochondrial dysfunction, oxidative stress, and vascular damage, among others.\cite{tumminia_type_2018, duarte_crosstalk_2013}

The connection between GLP-1RAs and improved cognition has been well-demonstrated in mouse models. Specifically, liraglutide (a GLP-1RA) reduced neuroinflammation and improved synaptic function (reducing amyloid formation).\cite{mcclean_diabetes_2011,mcclean_glucagon-like_2010} GLP-1RAs are thought to be particularly protective, both because of their effects on glucose regulation (and subsequently weight reduction) as well as a direct effect on neuroplasticity.\cite{duarte_crosstalk_2013,boye_generalizability_2019, onoviran_effects_2019} Secondary analyses of randomized placebo-controlled trial data in diabetic patients also showed a significant reduction in cognitive decline among those in the GLP-1RA treatment arm compared to the control arm.\cite{ballard_liraglutide_2020,vadini_liraglutide_2020,biessels_effect_2019} Due to the short-term follow-up of these randomized studies, the outcome of interest was either cognitive worsening (from validated cognition tests) or dementia-related adverse events. Studying dementia onset requires longer-term follow-up in large cohorts that are seldom practically or financially feasible for most randomized trials. Further, trial participants in these secondary analyses were also higher risk patients and not representative of the general population.

Observational studies circumvent these shortcomings by allowing access to expansive datasets with a sufficient number of events to look at relatively rare exposures or outcomes and evaluate long-term treatment effects. A growing body of observational research studying the effects of GLP-1RAs on dementia has shown a reduction in dementia risk.\cite{ballard_liraglutide_2020,norgaard_treatment_2022,wium-andersen_antidiabetic_2019} In the Danish National Registry specifically, one analysis showed a protective effect of GLP-1RA use on dementia for every year increase in exposure.\cite{norgaard_treatment_2022} While these results are encouraging, the application of the causal roadmap illustrates how they may fall short of estimating the causal effect of greatest interest. (1)\emph{Causal estimand specification}: some used a hazard ratio to measure effect\cite{ballard_liraglutide_2020,norgaard_treatment_2022}; while commonly used, the hazard ratio is not causally interpretable. A hazard is the instantaneous rate of having an event given that the event has not already occurred, and thus at each time point the hazard ratio compares event rates between study arms among participants who remain event-free. By definition, over time, this measure is susceptible to an inherent selection bias which can make it impossible to recover a counterfactual quantity generally of interest in a causal study.\cite{hernan_hazards_2010} (2)\emph{Identification}: in order to causally interpret results, we must satisfy the sequential randomization assumption (see Section 2.3.1), which states that the counterfactual outcome must be independent of treatment assignment, conditional on measured past. Adjusting for only baseline covariates may not be sufficient to adjust for confounding, particularly if there are time-dependent confounders that are affected by past treatment usage and affect future treatment usage.(3)\emph{Estimation}: estimation, particularly in the context of data sparsity and time-varying covariate histories, can prove challenging. The Cox model is is commonly used in longitudinal registry analyses. Estimators that can incorporate variable selection can optimize precision, and become particularly important in such cases. Integration of machine-learning for flexible adjustment can both reduce bias and improve precision, but must be incorporated into appropriate estimators.
%I like this paragraph, but not sure about this sentence. I think what you probably want to say is that prior studies used eg COx models. Integration of machine-learning for flexible adjustment can reduce both bias and precision, and is often essential for valid inference, but must be incorporated into an apprpriate estimator such as tatgeted maximum likleihood.

%For example, if the treatmentis detrimental early in follow-up, individuals who remain event-free in the%treatment arm will soon be at lower risk than individuals who remain event-freein the control arm. Susceptibility to the outcome is thus no longer balanced betweenarms; the treatment arm is instead at an inherent advantage.

%\subsection{Causal question}
Through the subsequent analysis, we apply the longitudinal causal roadmap to answer the question: what is the effect of sustained, cumulative exposure to GLP-1RAs vs i) any active comparator or, ii) specific active comparators (sodium-glucose cotransporter-2 or SGLT2 inhibitors, dipeptidyl peptidase 4 or DPP-4 inhibitors) on the cumulative risk of dementia by five years among patients with T2DM initiating second-line antihyperglycemic therapy for the first time? For each of these causal questions, we outline the causal estimand, specify the observed data and a causal model describing what is known about the real-world processes that generated it, and finally identify the causal estimand as a function of the observed data distribution that is equivalent to the causal estimand under explicit causal assumptions. This sets the stage for defining the statistical estimation problem and specifying an estimator, as we do in Section 3.

\subsection{Target causal estimand}

Our target population consists of insulin-naive Danish diabetes patients who initiate second-line diabetes medication for the first time. We follow these patients from index date $t=1$, corresponding to the date of first use of secondline diabetes medication, until a maximum of five year follow-up period, $t=1,...,K+1$, where each unit of time is six months in length.
The time $K+1$ is defined as the time at which the outcome (dementia) status is evaluated, in main analyses $K+1$ is set to five years, or ten 6-month long time intervals.
%\subsection{Treatment regimes of interest}
In this population, we are interested in two general causal questions, corresponding to two types of hypothetical interventions (also known as treatment strategies or treatment regimes). We are firstly interested in a hypothetical intervention only on GLP-1RA use throughout follow-up ($\bar{\tilde{A}_1}\equiv \tilde{A}_1(1),...,\tilde{A}_1(K+1))$, where $\tilde{A}_1(t)$ is an indicator that denotes use of GLP-1RA during time t, and where, here and throughout, an overbar denotes the the longitudinal history of a variable). Our hypothetical regimes of interest are sustained GLP-1RA use throughout follow-up (denoted by $\bar{\tilde{a}}_1=1$), and sustained GLP-1RA non-use throughout follow-up (denoted by $\bar{\tilde{a}}_1=0$); in both regimes we also hypothetically intervene to prevent censoring ($\bar{c}=0$, where $C(t)$ denotes an indicator of right censoring by time t). Let Y(t) denote an indicator that a participant is diagnosed with dementia by time t; for notational convenience, let $Y\equiv Y(K+1)$ denote diagnosis of dementia by the final time point of interest, $K+1$. Our counterfactual outcomes of interest under a hypothetical intervention on GLP-1RA use ($\bar{\tilde{A}_1}$) correspond to the diagnosis of dementia by time $K+1$ that would have been seen under these two hypothetical regimes (Table 1; these counterfactual quantities are formally defined using the causal model described Supplemental Materials 1): $Y_{\bar{\tilde{a}}_1=1,\bar{c}=0},Y_{\bar{\tilde{a}}_1=0,\bar{c}=0}$.
The above hypothetical regimes allow us to define our target causal estimand (or ``causal contrasts of interest''\cite{hernan_using_2016}). Here, we are interested in the cumulative causal risk difference for dementia diagnosis by five years if all patients had complied with the intervention arm vs. the control arm, intervening to prevent administrative censoring (Table 1): $E[Y_{\bar{\tilde{a}}_1=1, \bar{c}=0}]-E[Y_{\bar{\tilde{a}}_1=0, \bar{c}=0}]$.

We are secondly interested in a hypothetical study intervening on both exposure (GLP-1RA use, $\bar{\tilde{A}_1}$) and active comparator use ($\bar{\tilde{A}_2}$, where $\tilde{A}_2(t)$ denotes use of the active comparator, either SGLT-2 or DPP-4 use at time t). Specifically, we are interested in contrasting a regime of sustained GLP-1RA use throughout follow-up and no active comparator use throughout follow-up ($\bar{\tilde{a}}_1=1,\bar{\tilde{a}}_2=0$), with a regime of sustained non-use of GLP-1RAs throughout follow-up and sustained active comparator use throughout follow-up ($\bar{\tilde{a}}_1=0,\bar{\tilde{a}}_2=1$). In both regimes, we hypothetically intervene to prevent censoring ($\bar{c}=0$). 
%\subsection{Target causal parameters}
$E[Y_{\bar{\tilde{a}}_1=1,\bar{\tilde{a}}_2=0, \bar{c}=0}]-E[Y_{\bar{\tilde{a}}_1=0,\bar{\tilde{a}}_2=1, \bar{c}=0}]$, sustained GLP-1RA use, with no active comparator use and no administrative censoring throughout follow-up, compared to sustained active comparator use (e.g., SGLT2), with no GLP-1RA use and no administrative censoring throughout follow-up.

\subsection{Observed data and causal model}
The observed data available to estimate this casual estimand included all diabetes patients in the Danish National Registry who met the following inclusion criteria: were at least 50 years of age; had evidence of prior metformin use; initiated a second line medication between 2009 and 2021; were insulin-naive and dementia-free at index date (Supplemental Materials 2).

Data from this cohort can be described by a general longitudinal data structure. For ease of notation, we refer to the treatment and censoring processes together as  $A(t)=(\tilde{A}_1(t), \tilde{A}_2(t), C(t))$ and also refer to the time-varying covariate information $\tilde{L}(t)$, incident dementia diagnosis $Y(t)$, and death (the competing risk) $D(t)$ collectively as $L(t)=(\tilde{L}(t), Y(t), D(t))$. We denote the individual-level observed longitudinal data as:
$$ O = (W, L(1), A(1), ..., L(K), A(K), L(K+1)) \sim  P_0,$$
where W denotes baseline (time-invariant) covariate information. Note that $\tilde{A}(1)$ is exposure status (second-line diabetes medication use) at the beginning of the first node or time interval (i.e. index date), and $c(t=1)=0$ by definition. We follow patients through the registers until dementia, death, emigration, or August 2021, whatever comes first. 

Index date and baseline measurements are defined at the date of the start of second-line regimen. We use a discrete time scale with 6-month long time intervals. For notational convenience, we define variables after death, dementia diagnosis, or censoring as equal to their last observed value. For further explanation of the data-generating process, see Supplemental Materials 2.

%\subsubsection{Exposure}
We utilized prescription fill data from the Danish registry to define exposure to diabetes medications.  Specifically, a patient was defined as exposed to a medication at any time point if the patient redeemed at least one prescription of a medication in the corresponding time interval. For a full list of second-line diabetes medications used and relevant ATC codes, see Supplemental Materials 3.

%\subsubsection{Outcome Y(t): Dementia}
 We approximate our outcome, dementia onset, by time $t$ ($Y(t)$), through either of the following criteria:(1) dementia diagnosis, measured through inpatient diagnosis codes; or (2) purchase of dementia medication, measured through registry medication fill table. For a list of ICD and ATC codes used, see Supplemental Materials 3. Death ($D(t)$) is is treated as a competing risk. 
%\subsubsection{Covariates}
%For time-constant covariates, there was little to no missingness (\(\leq5\%\)), and we imputed to the median and mode for continuous and categorical covariates, respectively.
We consider the following baseline covariates: age (years), sex (male/female), education (basic, some college, college or higher), baseline income (tertiles), and time on metformin prior to index date (days). We also consider the following time-varying comorbidities: heart failure, renal disease, chronic pulmonary disease, any malignancy, ischemic heart disease, myocardial infarction, hypertension, and stroke. The time-varying comorbidities remain 0 until evidence of their presence through an inpatient ICD code in the registry is detected, then remain 1 for the remainder of follow up. We also include co-medication use, including: beta-blockers, calcium channel blockers, renin-angiotensin system-acting inhibitors, loop diuretics, Mineralocorticoid receptor antagonists and chronic obstructive pulmonary disease (COPD) medications. For a more detailed definition of all covariates, see Supplemental Materials 4. 

\subsubsection{Identification: causal to statistical estimand}

The translation of our causal estimand into a statistical estimand, i.e., a function of the distribution of the observed data, requires causal assumptions about the data-generating process beyond simply the time ordering of covariates. The core assumption in the assessment of long-term cumulative treatment effects is the sufficiency of measured variables to adjust for confounding of treatment decisions as well as informative censoring. Such an assumption can be assessed either from the casual graph directly using the sequential back door criteria, or stated in the language of counterfactuals as the sequential randomization assumption. The sequential randomization assumption\cite{laan_targeted_2018,robins_causal_1997} posits that at each time point, the counterfactual outcome $Y_{\bar{a}}$ is independent of treatment assignment and censoring, conditional on the history of measured covariates and prior treatment:
$$ Y_{\bar{a}} \perp \!\!\! \perp A(t) \mid  W, \bar{L}(t),\bar{A}(t-1)=\bar{a}(t-1), t=1,..,K,$$
for every treatment history $\bar{a}$ of interest. Of note, this could equivalently be written to more explicitly incorporate the distinct elements of $L(t)$, as
$$Y_{\bar{a}} \perp \!\!\! \perp A(t) \mid W, Y(t)=0, D(t)=0, \bar{\tilde{L}}(t),\bar{A}(t-1)=\bar{a}(t-1), t=1,..,K,$$
for every treatment history $\bar{a}$ of interest (see Table 1). These definitions are equivalent because for the other possible values of $\bar{Y}(t)$ and $\bar{D}(t)$ it holds by definition. 

For this assumption to hold, it is sufficient to assume in our causal model (Supplemental Materials 1), that we have no unmeasured common causes of our exposure and censoring nodes $A(t)$ at each time point and counterfactual outcome $Y_{\bar{a}}$. For each pairwise comparison, we consider other second-line medications not included in the contrast of interest as observable past.

We must also assume positivity---specifically, that there is a positive probability of continuing to follow the treatment of interest at each time point and no censoring, given previous treatment and covariate history\cite{robins_causal_1997, laan_targeted_2018}:
$$P(A(t)=a(t) \mid \bar{A}(t-1)=\bar{a}(t-1),\bar{L},W)>0, t=1,...,K  \hspace{.5cm}$$
for all possible covariate histories that can occur in the underlying data generating process among those who followed the regime. Threats to inference posed by violations (and near violations) of this assumption can be investigated empirically using the data, and used to inform specification of an estimator, as described further in the following section. 
%We empirically describe treatment support in the results (section 4.2) and in Supplemental Materials 5. 

Under these assumptions, the causal estimands corresponding to the counterfactual difference in risk of dementia by five years under sustained GLP-1RA use compared to sustained use of an active comparator, can be translated into a target statistical estimand using Robins' longitudinal g formula.\cite{robins_new_1986} In the present case study, we use the iterated conditional expectation form of this estimand.\cite{bang_doubly_2005}.

\section{Statistical model and estimator}

Prior to specifying a statistical estimator, the roadmap makes clear that it is crucial to specify a statistical model that incorporates only assumptions on the observed data distribution ($P_0$) that are known to hold. The existence of death without dementia as a competing risk provides an example of such knowledge in our case study: after death, the probability of incident dementia is deterministically zero. While such a statement may seem obvious, we illustrate below its explicit incorporation into the estimator. 

In addition to incorporating model knowledge, there are several decisions made in the estimation of the causal quantity of interest that require careful selection. In the following section, we highlight how simulations can be used to pre-specify an estimator and ensure its expected performance.

\subsection{Simulation}
Simulations are helpful tools that enable the selection of a valid and efficient estimation procedure. Simulations allow for the comparison of distinct estimation procedures. When registry data that are not always easily accessible due to privacy concerns, simulations also allow researchers to work with realistic longitudinal data remotely. We illustrate how simulations were used to pre-specify our statistical estimator, including the approach to inference. This includes the following choices: estimator choice (IPTW---commonly used in the literature, despite challenges---vs L-TMLE); approach to nuisance parameter estimation (specifically the pre-specification of machine learning algorithms to estimate the ``treatment mechanism''-–-conditional probabilities of medication use and censoring given the past---and iterated outcome regressions); and variance estimation (specifically, use of the empirical variance of the estimated  influence function, an alternative "robust" variance estimator\cite{tran_robust_2023}, or the non-parametric bootstrap).

\subsubsection{Data generating process}
To simulate data with the complex longitudinal structure found in the Danish registry, we ran logistic regression models on the real registry cohort of interest (Section 5.1.1). We used these models to estimate each node at each time point, conditional on the observable past. We then created a matrix of beta coefficients from each of these regressions and exported them from the Danish server.

Using these coefficients, we simulated n=500 datasets which contain follow-up data of 100,000 simulated patients. We simulated a first scenario (scenario 1) with no relationship between exposure and outcome, and a second scenario (scenario 2) with a significant relationship between exposure and outcome. We simulated the datasets using the lava package.\cite{holst_linear_2013} For each node at each time point, we used the vector of beta coefficients to predict the node values. This dataset has a dependent correlation structure between exposure and outcome, reflective of the parametric relationship fitted in the observed data. We then permuted the outcome, competing risk and censoring nodes in the dataset to create the scenario where the true risk difference between the treatment regimens is zero. This allowed us to approximate the complex correlation structure of treatment and covariate measures in the data.

We calculated the true causal risk difference between full GLP-1RA use throughout follow-up and no GLP-1RA use throughout follow-up in the first simulation scenario by simulating large datasets with counterfactual covariates and outcomes setting all exposure nodes ($\tilde{A}_1, \tilde{A}_2$) to 1 or 0, depending on the causal contrast of interest. For further detail, see Supplemental Materials 5.

\subsubsection{Estimation procedure}

We considered both IPTW estimator and L-TMLE as candidate estimators in the simulation. L-TMLE is a double-robust plug-in estimator that enables to control for time-dependent confounding without adjusting away the effect of the exposure on the outcome through downstream mediators. \cite{laan_targeted_2011,laan_targeted_2018} Nuisance parameters in L-TMLE estimation can incorporate machine learning algorithms while maintaining statistically valid inference under assumptions, including, informally, an assumption that initial estimators of the outcome regressions are not overfit.\cite{laan_targeted_2011,laan_targeted_2018} While the use of internal sample splits and cross-fitting approaches (such as cv-TMLE\cite{laan_targeted_2018,zheng_asymptotic_2010}) can partially address these challenges, the lack of a computationally efficient implementation of the cv-TMLE estimator for the longitudinal treatment effect of interest led us to instead rely on simulations to carefully pre-specify the approach to nuisance parameter estimation. 

We considered the following candidate algorithms for treatment propensity and censoring mechanisms and iterated conditional outcome regression estimation: logistic regression with LASSO penalty\cite{friedman_regularization_2010,tay_elastic_2023} (one with cross-validated $\lambda$ hyperparameter---the $\lambda$ value with minimum mean cross-validated error---and one undersmoothed---choosing the minimum penalization across a range of candidate $\lambda$ values), logistic regression with ridge and elastic net\cite{friedman_regularization_2010,tay_elastic_2023} penalty (one with $\lambda$ chosen to minimize CV error, and one undersmoothed), unpenalized logistic regression model (identity link, adjusted and unadjusted), and random forest\cite{wright_ranger_2017}. We applied all candidate estimators and estimation procedures to both scenario 1 and secnario 2 data generating processes. 
%To estimate variance, we considered empirical influence curve variance, “robust” variance estimator \cite{tran_robust_2023}, and non-parametric bootstrap. To handle competing events, commonly, this information is incorporated into a censoring node. In our case, we are interested in death as a competing event. We examined estimates both with and without a deterministic function. Using a deterministic function,  we can set Q regression probabilities to 0 once a competing event, like death, has occurred.
Finally, prior work demonstrates that even when a point estimator performs well (in that its bias to standard error ratio is low enough to allow for valid statistical inference, implying nominal ``oracle coverage’’ i.e. confidence interval coverage treating the true variance as known rather than estimated, calculated using the empirical variance across simulation repetitions), standard approaches to variance estimation based on the empirical variance of the estimated influence function, commonly used for both the IPTW and TMLE estimators, may underestimate the true variance, particularly in challenging settings with rare outcomes and/or practical positivity violations, leading to under-coverage.\cite{tran_robust_2023} Alternative approaches to variance estimation include a non-parametric bootstrap, and an approach based on fitting a separate TMLE of the variance itself, implemented as the ``robust'' option in the ltmle package\cite{ tran_robust_2023}. As each of these approaches has its own vulnerabilities, we further employed simulations to pre-specify an approach to variance estimation and inference.

As described previously, the incorporation of death as a competing event for dementia onset implies knowledge about the statistical model; namely, that once a death occurs, the probability of dementia is deterministically zero. By incorporating this knowledge in the estimation of the iterated outcome regression, the model will be correctly specified (i.e. will incorporate essential subject matter knowledge). We therefore examined all candidate estimators and algorithms described above with this knowledge explicitly incorporated (using the deterministic Q function in the \textit{ltmle} R package).\cite{gruber_tmle_2012} Code used for the simulation is available in an online repository.\cite{mertens_registry_simulations_2023}

\subsection{Simulation and estimation selection}

We evaluated performance under each combination of the above specifications on 500 simulated datasets generated from our data-generating process described above. To evaluate the performance of the estimation procedures, we compared bias, variance, bias to standard error ratio, 95 \% confidence interval coverage, and oracle coverage (Supplemental Materials 6).

%) both scenarios-  TMLEs had oracle coverage closer to nominal  than IPTW  (only accurate if cut the RR). so foucus rest of discussion on TMLEs.  AMong the TMLE estimators, 
%2) among the TMLEs using ML,
%for senario 1,  penalized regression estimators with undersmoothing and truncation had  the lowest variance among estimators with reasonable oracle coverage. 
%For scenrio 2, penalize dregressions with CV min SE also had close to nomoinal CI coverage and  a bit lower variance, although fairly comparable.  However there oracle coverage was anticonservative in scenario 1, so did not choose them.
%3) GLM with truncation also low variance and close ot nominal coverage (likely an artifact of thow the data were simulated)- given that the truncate undermoothed penalized regression based estimators performed similarly across both secenrios and provide more felxibility, went with that approach. 
%4).among these, chosee  ridge because...??? (not actually sure- it is higher variance than the other penalized regression estimators given close to nominal CI in both scenrios...I would have gone with elestic net or lasso based on the results you show- that said not worth holding up submission based on this but nice if you could come up with a reason...?
%5) Once had selected that point esitmator, you evaluated approaches to variance estimation and you found closest to nominal coverage using the np bootstrap. (Maybe only for scenario 1? in which case need ot say that)

In both scenario 1 and 2, we found the TMLE estimator to have nominal oracle coverage (i.e. closer to 95\%), compared to the IPTW estimator. Among the TMLE candidate algorithms for nuisance parameter estimation (LASSO, Ridge, Elastic Net, Random Forest) in Scenario 1, we found the undersmoothed penalized regression algorithms with g-truncation performed best---i.e. had nominal oracle coverage and lowest variance (Supplemental Materials 5 and 6). We then compared variance estimation methods; we evaluated variance estimators for Scenario 1 only due to computational expense. The non-parametric bootstrap was the variance estimation option with the coverage closest to 95\%; we found TMLE variance estimation was conservative and influence-curve-based estimation was anti-conservative.  

Based on simulation results for the risk difference in scenario 1 (the realistic data generating process), we chose the L-TMLE estimator using undersmoothed ridge regression with g-truncation. This estimation strategy had the best oracle coverage, as well as low bias and variance. For more detailed results, see Supplemental Materials 6.

%\subsection{Analysis}
%We then implemented the final analysis on the registry data. We additionally examined several pre-specified sensitivity analyses. We implemented the analysis using longer (12-month) and shorter (3-month) time intervals in order to assess the robustness of the results under differing levels of granularity. We restricted the data to later start years (2014 and later); this was to accommodate both the slower adoption of GLP-1RAs and the later start of SGLT2 inhibitors, which came to market in 2014.

For all analyses, we used R version 4.2\cite{r_package}, and the \textit{ltmle} package\cite{gruber_tmle_2012}.

\section{Results}

\subsection{Cohort demographics}

Of the registered diabetes patients using metformin in the Danish registry and initiating second-line treatment between 2009 and 2021 (n=154,128), 104,928 met inclusion criteria (Figure 1).

Among those meeting inclusion criteria for the study, 13\% (n=13,334) initiated GLP-1RAs at baseline, and 87\% (n=91,594) initiated other second-line medications (for further detail on demographics of other treatment medication initiators, see Supplemental Materials 7). Those who initiated GLP-1RAs at index date were more likely to be below 55, to be more educated, have a higher income, and have a longer duration of diabetes (Table 2). Those initiating GLP-1RAs at baseline were more likely to have some conditions (chronic pulmonary disease), and less likely to have others (heart failure, renal disease, ischemic heart disease, myocardial infarction). Those initiating GLP-1RAs were more likely to be treated with Mineralocorticoid receptor antagonists (MRAs), Calcium channel blockers (CCBs), or Renin–angiotensin system inhibitors (RASIs).

%By the end of the 5-year follow-up, n=46,447 were censored (44\%). More patients were administratively censored in the treatment than the control group(70\% vs 41\%), as expected due to date at which GLP1-1RA became widely available. During the patient followup time we observed n=27 cases of dementia among those who initiated GLP-1RA at baseline, and n=1,819 cases of dementia among those initiating another second-line drug at baseline.

\subsection{Risk differences}

We applied our pre-specified estimator, LTMLE, with propensity scores estimated using undersmoothed ridge regression,%LASSO (CV-lambda), 
and inference based on the non-paramteric bootstrap, to the registry data. The results show a protective effect comparing five-year sustained GLP-1RA use to sustained non-use (risk difference percent RD\% -0.48, 95\% CI -0.94,-0.01; Table 3) and to sustained use of DPP4 inhibitors (RD\% -0.48, 95\% CI -0.93,-0.02; Table 3). No significant differences were found comparing GLP-1RAs to SGLT2 inhibitors (RD\% -0.48, 95\% CI -0.93,-0.02; Table 3). In sensitivity analyses, we varied the time interval length, and found these results to hold irrespective of time discretization (Supplemental Materials 8); shortening the time unit led to a wider confidence interval. (We include relative risk results in Supplemental Materials 9.)

We additionally plotted the marginal risk of dementia at each six-month interval, among those who have survived until that time point and have followed the longitudinal treatment regimes of interest (sustained GLP-1RA use, sustained non-use) up until that time point. Results show a difference in the risk of dementia over time between the longitudinal exposure groups (Figure 2; Table 4).

Figure 2: The L-TMLE dementia risk estimates comparing sustained use and non-use of GLP-1RAs in a cohort of Danish second-line diabetes drug initiators 2009-2021.

\section{Discussion}

In this case study, we apply the casual roadmap to a longitudinal analysis of a large national registry database. Through the use of the roadmap, we explicitly state our causal estimand and the assumptions required to be able to estimate this quantity from the observed data. We found a significant relationship between sustained GLP-1RA use and five-year dementia risk compared to no usage of GLP-1RAs and compared to sustained DPP4 use, after adjusting for a wide range of time invariant and time-varying confounders using a pre-specified L-TMLE esitmator that incorporates machine learning. Effects of GLP-1RAs compared to active comparator SGLT2 inhibitors were not significant; however, confidence intervals were wide due to rare continuous usage of SGLT2 and greater administrative censoring as SGLT2 is a newer drug (usage increasing after 2011). High administrative censoring of the treatment group is likely due to the later introduction and relatively slower adoption of GLP-1RAs relative to other, more established second-line therapies.\cite{persson_different_2018} Our findings agree with prior literature on the benefit of GLP-1RAs on dementia onset in the Danish registry\cite{norgaard_treatment_2022}, observational studies,\cite{onoviran_effects_2019,wium-andersen_antidiabetic_2019} and some randomized trials\cite{ballard_liraglutide_2020,vadini_liraglutide_2020,cukierman-yaffe_effect_2020,gejl_alzheimers_2016}. 

As with all studies, ours is not without limitations. Firstly, due to data unavailability we were not able to adjust for smoking status or body mass index, which may have resulted in residual confounding of our effect estimate---leading to either under or overestimates of the protective effect of continuous GLP1-RAs. In addition, while this sample is generalizable nationally to Denmark and to other Nordic countries, the ethnically homogeneous population in the Danish registry does not necessarily generalize to more diverse populations. Finally, the data from our simulation was generated using coefficients from sequential regression models; this imposes a simpler parametric structure on the data generating process than may exist in reality. This simplified structure may lead to some overestimation of model performance, as some of the complexities of the true data distribution are not captured.

We found the implementation of a complex estimator such as L-TMLE with machine learning at the scale national registry to be feasible. However, implementation of such an analysis involves a variety of estimation decisions. Simulation provided a means to pre-specify these decisions and evaluate the performance of the resulting estimator in data generating processes designed to resemble the data, including to evaluate the performance (precision and confidence interval coverage) in the presence of rare events and long-term follow up. The simulations also provided additional evidence that if implemented with care, a TMLE estimator with machine learning can reduce bias and variance and improve inference relative to both IPTW estimators and TMLE estimators implemented without machine learning. Nonetheless, in this setting or rare outcomes and practical postivity violations, variance estimation remained a challenge. A common approach used for IPTW,  TMLE, and other estimators, based on the empirical variances of the estimated influence curve, resulted in markedly anti-conservative confidence interval coverage, as has been reported by others.\cite{tran_robust_2023} While the non-parametric bootstrap provided coverage closer to nominal it was still slightly anti-conservative.  Future work is needed to further improve variance estimation for complex longitudinal estimators in settings such as these.

While simulation results presented are interesting in their own right, the results must be interpreted locally to this particularly analysis. We recommend that investigators construct their own simulations with data similar to their own in key ways, rather than follow the estimation specifications here or use estimation defaults. To our knowledge, this is the first study of its kind of apply L-TMLE using longitudinal causal roadmap \cite{petersen_causal_2014, petersen_applying_2014} to a large national registry database.

%\subsection*{Acknowledgements}
%We are grateful for the help of our colleagues in the Joint Initiative for Causal Inference---a joint collaboration between UC Berkeley, University of Copenhagen, and Novo Nordisk. Their feedback has been integral to the manuscript.

%\subsection*{Author contributions}
\subsection*{Disclosures}

NN reports tuition and stipend support from a philanthropic gift from the Novo Nordisk
corporation to the University of California, Berkeley to support the Joint Initiative for
Causal Inference. 
AM and ZW have received salary compensation from the same philanthropic gift from Novo Nordisk.
MvdL reports that he is a co-founder of the statistical software start-up company TLrevolution, Inc. 
MvdL, MP, TG, CTP, TL and BZ report personal compensation for consultation from Novo Nordisk.
KK is employed by Novo Nordisk A/S and own stocks in Novo Nordisk A/S.

\subsection*{Funding information}
This work was funded through a philanthropic donation from Novo Nordisk.
\clearpage 
\printbibliography

\newpage
\begin{landscape}

\begin{table}[H]

%\begin{table}
\begin{center}

%\noindent\setlength\tabcolsep{4pt}%
%\begin{tabularx}{\linewidth}{l|c*{3}{>{\RaggedRight\arraybackslash}X}}
 %\begin{tabularx}{1.5\textwidth}{X*{4}{>{\RaggedRight}X}} 
\begin{tabular}{p{3cm}|p{5cm}|p{5cm}|p{5cm}}
\hline
& Causal estimand 1: 
\newline
  GLP1 vs. no GLP1& Causal estimand 2: 
  \newline
  GLP1 vs. SGLT2&Causal estimand 3: 
  \newline
  GLP1 vs. DPP4\\
  \hline
  Eligibility criteria &
  Dementia and insulin-naive diabetes patients initiating second-line therapies between 2009-2021 &
  Dementia and insulin-naive diabetes patients initiating second-line therapies between 2014-2021&
  Dementia and insulin-naive diabetes patients initiating second-line therapies between 2009-2021\\ 
  \hline
  Treatment regime & \vspace{1mm}
  Sustained GLP1 use ($\bar{\tilde{a}}_1=1$) throughout follow up, intervening to prevent censoring  &
  Sustained GLP1 use ($\bar{\tilde{a}}_1=1$) and no SGLT2 use ($\bar{\tilde{a}}_2=0$) throughout followup, intervening to prevent censoring&
  Sustained GLP1 use ($\bar{\tilde{a}}_1=1$) and no DPP4 use ($\bar{\tilde{a}}_2=0$) throughout followup, intervening to prevent censoring\\
  \hline
  Counterfactual outcome under treatment&

  $$Y_{\bar{\tilde{a}}_1=1, \bar{c}=0}(t)$$&
  \multicolumn{2}{c}{
  $Y_{\bar{\tilde{a}}_1=1, \bar{c}=0}$
  } \\
  \hline
  Control regime &
  No GLP1 use ($\bar{\tilde{a}}_1=0$) throughout follow-up, intervening to prevent censoring&
  Sustained SGLT2 use ($\bar{\tilde{a}}_2=1$) and no GLP1 use ($\bar{\tilde{a}}_1=0$) throughout follow up, intervening to prevent censoring&
  Sustained DPP4 use ($\bar{\tilde{a}}_2=1$) and no GLP1 use ($\bar{\tilde{a}}_1=0$) throughout follow up, intervening to prevent censoring\\ 
  \hline
  Counterfactual outcome under control &
   $Y_{\bar{\tilde{a}}_1=0, \bar{c}=0}(t)$&
 \multicolumn{2}{c}{
 $Y_{\bar{\tilde{a}}_1=0,\bar{\tilde{a}}_2=1, \bar{c}=0}(t)$
 }\\ 

 \hline
 Target causal parameter &
  $E[Y_{\bar{\tilde{a}}_1=1, \bar{c}=0}]-E[Y_{\bar{\tilde{a}}_1=0, \bar{c}=0}]$
&\multicolumn{2}{c}{
$E[Y_{\bar{\tilde{a}}_1=1,\bar{\tilde{a}}_2=0, \bar{c}=0}]-E[Y_{\bar{\tilde{a}}_1=0,\bar{\tilde{a}}_2=1, \bar{c}=0}]$
}\\
  \hline
%\end{tabularx}
\end{tabular}
\captionof{table}{Causal estimands for three key target comparisons of second line medication use}

\end{center}

%\end{table}

\end{table}
\newpage

%\begin{center}
%\begin{tabular}{llllll}
\captionof{table}{Baseline demographic characteristics of a cohort of diabetes patients initiating second line treatment for the first time between 2009 and 2021 in the Danish registry, stratified by GLP1-RA initiation at baseline. N(\%)}
%\begin{center}
 %\tabcolsep=1pt\relax
% \small% <<--- CAN USE \footnotesize OR \scriptsize IF NEEDED
%\makebox[\textwidth]
\begin{tabular}{ l| c c c c c } 
%\begin{adjustwidth}{-1in}{-1in}% adjust the L and R margins by 1 inch 
\noindent\setlength\tabcolsep{4pt}%
%\begin{tabularx}{\linewidth}{l|c*{4}{>{\RaggedRight\arraybackslash}X}}
%\begin{tabularx}
%  Variable&Level&GLP1RA \newline(n=13335)&no GLP1RA \newline(n=91598)&Total \newline(n=104933)\\
  %&p-value \\
%  \hline

Variable & Level & GLP1-RA \newline (n=13,334) & no GLP1-RA \newline (n=91,594) & Total \newline (n=104,928) \\%& p-value\\
\hline
Age category & $<$ 55 & 2,851 (21.4) & 12,106 (13.2) & 14,957 (14.3) \\%& \\
 & 55-60 & 2,925 (21.9) & 14,348 (15.7) & 17,273 (16.5) \\%& \\
 & 60-65 & 2,583 (19.4) & 15,942 (17.4) & 18,525 (17.7) \\%& \\
 & 65-70 & 2,166 (16.2) & 15,834 (17.3) & 18,000 (17.2) \\%& \\
 & 70-75 & 1,557 (11.7) & 14,078 (15.4) & 15,635 (14.9) \\%& \\
 & 75-80 & 821 (6.2) & 9,980 (10.9) & 10,801 (10.3) \\%& \\
 & 80-85 & 306 (2.3) & 5,823 (6.4) & 6,129 (5.8) \\%& \\
 & 85-90 & 92 (0.7) & 2,528 (2.8) & 2,620 (2.5) \\%& \\
 & $>$ 90 & 33 (0.2) & 955 (1.0) & 988 (0.9) \\%& < 1e-04\\
\hline
Education & Basic & 4,634 (34.8) & 38,764 (42.3) & 43,398 (41.4) \\%& \\
 & Medium & 6,471 (48.5) & 41,043 (44.8) & 47,514 (45.3) \\%& \\
 & High & 2,229 (16.7) & 11,787 (12.9) & 14,016 (13.4) \\%& < 1e-04\\
\hline
Income (tertile) & 1 & 3,103 (23.3) & 31,923 (34.9) & 35,026 (33.4) \\%& \\
 & 2 & 3,943 (29.6) & 31,008 (33.9) & 34,951 (33.3) \\%& \\
 & 3 & 6,288 (47.2) & 28,663 (31.3) & 34,951 (33.3) \\%& < 1e-04\\
\hline
Diabetes duration (years) & $<$ 1 & 3,747 (28.1) & 31,795 (34.7) & 35,542 (33.9) \\%& \\
 & 1-5 & 5,379 (40.3) & 33,287 (36.3) & 38,666 (36.9) \\%& \\
 & 5-10 & 3,082 (23.1) & 20,932 (22.9) & 24,014 (22.9) \\%& \\
 & $>$ 10 & 1,126 (8.4) & 5,580 (6.1) & 6,706 (6.4) \\%& < 1e-04\\
\hline
  Diagnoses &&\\
\hspace{3mm}Heart failure &   & 614 (4.6) & 5,886 (6.4) & 6,500 (6.2) \\%& < 1e-04\\
\hspace{3mm}Renal disease &   & 91 (0.7) & 1,134 (1.2) & 1,225 (1.2) \\%& < 1e-04\\
\hspace{3mm}Chronic pulmonary disease &   & 906 (6.8) & 5,795 (6.3) & 6,701 (6.4)\\% & 0.040827\\
\hspace{3mm}Any malignancy &   & 689 (5.2) & 6,366 (7.0) & 7,055 (6.7) \\%& < 1e-04\\
\hspace{3mm}Ischemic heart disease &   & 1,416 (10.6) & 11,310 (12.3) & 12,726 (12.1) \\%& < 1e-04\\
\hspace{3mm}Myocardial infarction &   & 675 (5.1) & 5,149 (5.6) & 5,824 (5.6)\\% & 0.008918\\
\hspace{3mm}Hypertension &   & 2,876 (21.6) & 20,249 (22.1) & 23,125 (22.0)\\% & 0.164468\\
\hspace{3mm}Stroke &   & 454 (3.4) & 4,126 (4.5) & 4,580 (4.4) \\%& < 1e-04\\
\hline
Medications &&\\
   \hspace{3mm}Beta-blockers &   & 3,655 (27.4) & 27,185 (29.7) & 30,840 (29.4) \\%& < 1e-04\\
 \hspace{3mm}Calcium channel blockers &   & 4,107 (30.8) & 27,105 (29.6) & 31,212 (29.7) \\%& 0.004487\\
 \hspace{3mm}Renin-angiotensin system-
\\
\hspace{3mm}acting inhibitors   &   & 8,710 (65.3) & 57,552 (62.8) & 66,262 (63.1) \\%& < 1e-04\\
%thiazid &   & 2338 (17.5) & 14499 (15.8) & 16,837 (16.0) \\%& < 1e-04\\
 \hspace{3mm}Loop diuretics &   & 2,112 (15.8) & 14,538 (15.9) & 16,650 (15.9) \\%& 0.932430\\
 \hspace{3mm}Mineralocorticoid receptor
 \\\hspace{3mm} antagonists &   & 897 (6.7) & 5,495 (6.0) & 6,392 (6.1) \\%& 0.001099\\
\hspace{3mm}COPD medications &   & 2,108 (15.8) & 11,781 (12.9) & 13,889 (13.2) \\%& < 1e-04\\
%\end{tabular}
%\end{center}
%\end{tabularx}

%\end{center}
%\end{adjustwidth}
\end{tabular}
\clearpage
% \input{table3_sept2023}
% \clearpage
% \begin{flushleft}
% %\begin{tabular}
% \input{table2_new}
%\end{tabular}
% \end{flushleft}
\restoregeometry

%%%%%%%%%%%%%%%FIGURES%%%%%%%%%%%%%%%

 \begin{figure}[H]
 \includegraphics[width=\textwidth]{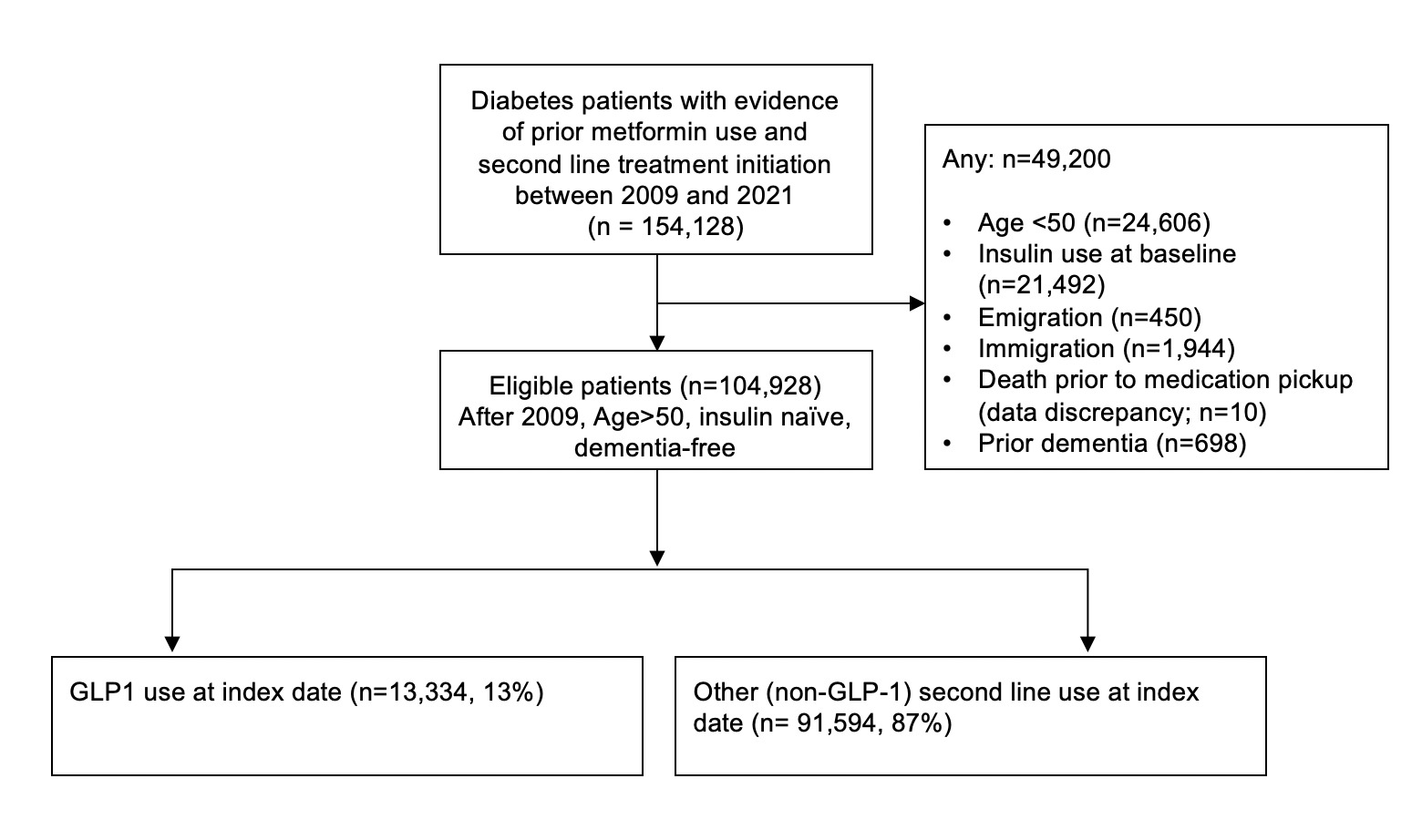}
 \centering
 \caption{Flowchart of selection criteria for a retrospective cohort of diabetes patients initiating second-line therapy for the first time between 2009 and 2021 in Denmark}
 \end{figure}

\newpage
 \begin{flushleft}
%\newgeometry{left=3cm,bottom=0.1cm}
\begin{figure}[H]
\includegraphics[width=0.7\textwidth]{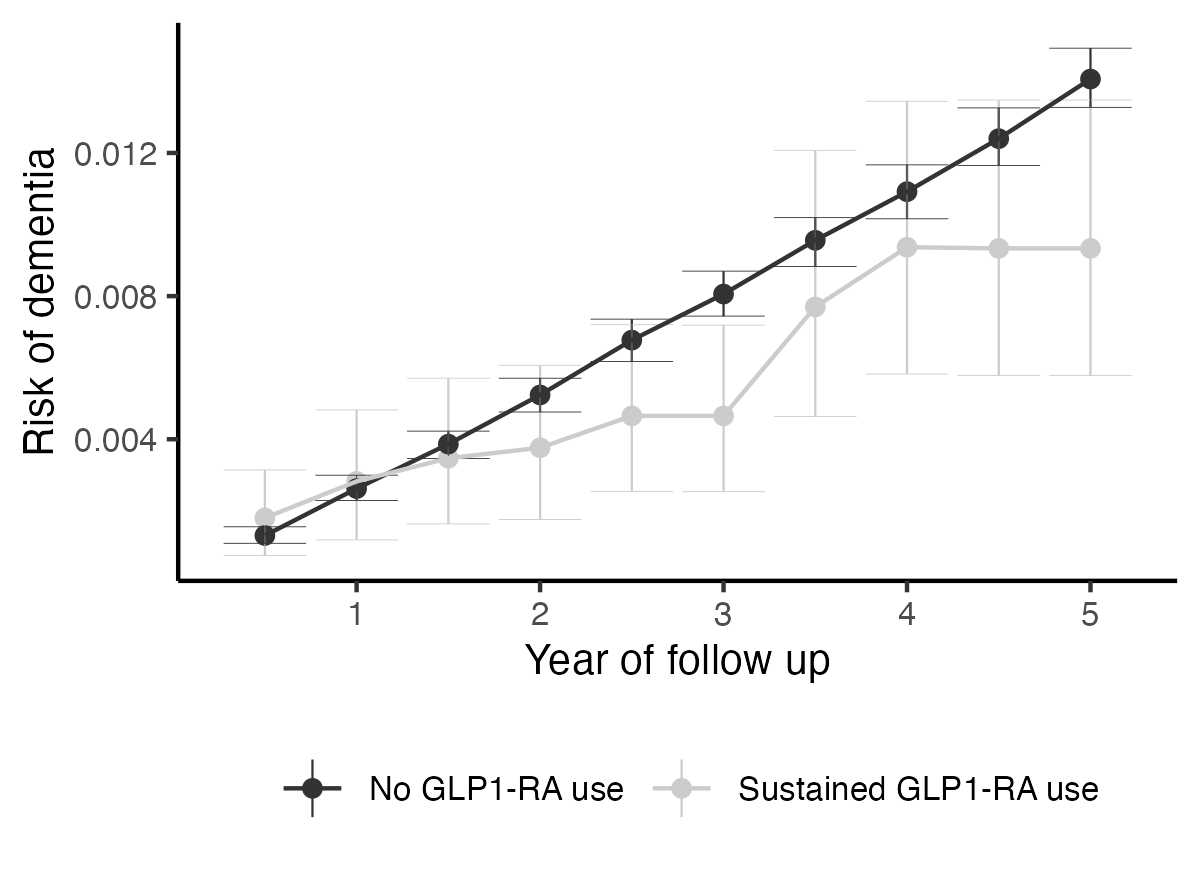}
\centering
\caption{The L-TMLE dementia risk estimates comparing sustained use and non-use of GLP-1RAs in a cohort of Danish second-line diabetes drug initiators 2009-2021.}
\end{figure}
%\begin{tabular}

\begin{table}[!ht]
    \centering
    \begin{tabular}{l|lllllllllll}
    \hline
            Interval & 0 & 1 & 2 & 3 & 4 & 5 & 6 & 7 & 8 & 9 & 10 \\ \hline

 GLP1-RA use & 13,334 & 10,419 & 7,627 & 6,008 & 4,983 & 4,282 & 3,587 & 3,084 & 2,746 & 2,484 & 2,216 \\ \hline
        \hspace{3mm}Death & 0 & 96 & 82 & 63 & 59 & 46 & 40 & 36 & 47 & 43 & 45 \\ %\hline
        \hspace{3mm}Dementia & 0 & 9 & $<$5* & $<$5* & $<$5* & $<$5* & $<$5* & 5 & $<$5* & $<$5* & $<$5* \\ %\hline
        \hspace{3mm}End of follow-up & 0 & 1,992 & 2,052 & 1,456 & 995 & 686 & 714 & 468 & 357 & 269 & 290 \\ %\hline
        \hspace{3mm}Non-adherent & 0 & 818 & 1,473 & 1,571 & 1,540 & 1,507 & 1,447 & 1,441 & 1,372 & 1,321 & 1,253 \\ %\hline
            \hline
        No GLP1-RA use & 91,594 & 82,666 & 74,352 & 67,744 & 61,901 & 56,768 & 51,814 & 47,492 & 43,221 & 39,505 & 35,552 \\ \hline
        \hspace{3mm}Death & 0 & 1,740 & 1,361 & 1,243 & 1,137 & 1,094 & 977 & 967 & 925 & 876 & 814 \\ %\hline
        \hspace{3mm}Dementia & 0 & 114 & 111 & 99 & 94 & 107 & 88 & 85 & 78 & 79 & 79 \\ %\hline
        \hspace{3mm}End of follow-up & 0 & 4,525 & 5,126 & 4,039 & 3,871 & 3,327 & 3,688 & 3,078 & 3,456 & 2,911 & 3,277 \\ %\hline
        \hspace{3mm}Non-adherent & 0 & 2,549 & 4,265 & 5,492 & 6,233 & 6,838 & 7,039 & 7,231 & 7,043 & 6,893 & 6,676 \\ %\hline

    \end{tabular}
    \captionof{table}{Table of longitudinal patterns in treatment and control groups in a cohort of diabetes patients initiating second line use between 2009 and 2021 in Denmark \newline
    \hspace{3mm}*cells with less than five observations are not shown in the interest health information privacy}
\end{table}

%\end{tabular}
 \end{flushleft}
% \restoregeometry
\end{landscape}
\newpage

\begin{table}[!ht]
\centering
\begin{tabular}{>{\hspace{0pt}}m{0.3\linewidth}|>{\hspace{0pt}}m{0.2\linewidth}>{\hspace{0pt}}m{0.2\linewidth}>{\hspace{0pt}}m{0.2\linewidth}}

%\begin{tabularx}{l|c*3}
Comparison & Treatment Risk & Control Risk  &  Risk Difference \% \newline (RD\%, 95\% CI) \\
\hline
1) GLP1 vs. no GLP1 & 0.0093  &0.0141  & -0.47(-0.94,-0.01 )  \\
2) GLP1 vs. SGLT2 & 0.0071 & 0.0074 & -0.03 (-0.27,0.22) \\
3) GLP1 vs. DPP4 & 0.0068 & 0.0115 & -0.48 (-0.93,-0.02) 
\end{tabular}
\caption{Five-year risk of dementia under sustained GLP1-RA use vs sustained comparator use in a cohort of Danish diabetes patients initiating second-line therapy (RD represented as \%, rescaled x100)}
\end{table}

%\end{tabular}
%\end{center}



\section*{Supplementary material}

# Supplemental Materials

# Contents

# 1 Structural causal model

We define the following exogenous nodes for $t = 1...k + 1$:

$$X = W, L(t), A(t)$$

where: $W$ includes all time-constant covariates: age (years), sex (male/female), education (basic, some college, college or higher), baseline income (tertiles), time on metformin prior to index date (days).

$$A(t) = (\tilde{A}_1(t), \tilde{A}_2(t), C(t))$$

- $\tilde{A}_1(t)$ is GLP1-RA use at time t
- $\tilde{A}_2(t)$ is active comparator use (however designated) at time t
- $C(t)$ is administrative censoring at time t

$$L(t) = (\tilde{L}(t), Y(t), D(t))$$

- $\tilde{L}(t)$ is the time-varying covariate status at time t, these include diagnoses (heart failure, renal disease, chronic pulmonary disease, any malignancy, ischemic heart disease, myocardial infarction, hypertension, stroke) and medication use (beta-blockers, calcium channel blockers, renin-angiotensin system-acting inhibitors (RASI), loop diuretics, Mineralocorticoid receptor antagonists (MRAs) and COPD medications)
- $Y(t)$ is the outcome, dementia diagnosis, at time t
- $D(t)$ is death at time t

And exogenous error:

$$U = (U_A, U_L)$$

Structural equations are as follows:

$$L(t) = f_{L(t)}(\bar{A}(t-1), \bar{L}(t-1), U_{L(t)}), t = 1, ..., K+1$$

$$A(t) = f_{A(t)}(\bar{A}(t-1), \bar{L}(t-1), U_{A(t)}), t = 1, ..., K$$

# 2 Cohort selection and data processing

We collected a retrospective cohort of diabetes patients who:

- were at least 50 years of age
- had at least one medication fill for metformin use (Table 1)
- did not immigrate to Denmark in the five years prior to beginning a second line therapy or emigrate before the start of the study
- initiated a second line medication (Table 2) between January 1st, 2009 and July 20th, 2022 (second line start date as the index date)
- were insulin-naive at index date (Table 1)
- were dementia-free at index date (Table 1)

Table 1: List of codes used for inclusion and exclusion criteria in cohort of second line diabetes drug initiators

| drug | ATC |
| --- | --- |
| Insulin | $A10A$ |
| Metformin | $A10BA02, A10BD02, A10BD03, A10BD05, A10BD07, A10BD08,$ $A10BD10, A10BD11, A10BD13, A10BD14, A10BD15, A10BD16,$ $A10BD17, A10BD18, A10BD20, A10BD22, A10BD23, A10BD25,$ $A10BD26$ |

Table 2: List of codes used to define second line diabetes medication use, for inclusion and exposure node definition

| Node | Drug | ATC Codes |
|---|---|---|
| $\tilde{A}_1$ | GLP1-RA | A10BJ |
| $\tilde{A}_2$ | SGLT2 Inhibitors | A10BK, A10BD15, A10BD16, A10BD19, A10BD20, A10BD21, A10BD23, A10BD24, A10BD25 |
| | Sulfonylureas | A10BB, A10BD01, A10BD02, A10BD04, A10BD06 |
| | DPP-4 Inhibitors | A10BH, A10BD07, A10BD08, A10BD09, A10BD10, A10BD11, A10BD12, A10BD13, A10BD18, A10BD19, A10BD21, A10BD22, A10BD24, A10BD25 |
| | Thiazolidinediones | A10BG, A10BD03, A10BD04, A10BD05, A10BD06, A10BD09, A10BD12, A10BD26 |
| | Sulfonamides | A10BC |
| | Alfa-glucosidase Inhibitors | A10BF, A10BD17 |
| | Repaglinides | A10BX, A10BD14 |
| | Biguanides | A10BA01, A10BA03, A10BD01 |

# 3 Treatment and outcome definitions

**Treatment exposure:** Usage of each second-line medication (Table 2) was defined for each time node (6-month bins in main analyses) if the medication coverage time (medication pick up date +6 months) overlapped with the time node. **Outcome:** Dementia onset date was defined as the first occurrence of either (1) a dementia medication pickup, or (2) a dementia diagnosis during follow up. (Table 3)

Table 3: ATC and ICD codes used to define dementia endpoint

| Code | Type | Description |
| --- | --- | --- |
| N06DA01 | ATC | Tacrine |
| N06DA02 | ATC | Donepezil |
| N06DA03 | ATC | Rivastigmine |
| N06DA04 | ATC | Galantamine |
| N06DA05 | ATC | Ipidacrine |
| N06DA52 | ATC | Donepezil and memantine |
| N06DA53 | ATC | Donepezil, memantine and Ginkgo folium |
| 290 | ICD-08 | Dementias |
| F00 | ICD-10 | Dementia in Alzheimer disease |
| F01 | ICD-10 | Vascular dementia |
| F02 | ICD-10 | Dementia in other diseases classified elsewhere without behavioral disturbance |
| F03 | ICD-10 | Unspecified dementia |
| G30 | ICD-10 | Alzheimer's disease |
| G31.1 | ICD-10 | Senile degeneration of brain, not elsewhere classified |
| F051 | ICD-10 | Delirium superimposed on dementia |
| 331 | ICD-08 | Alzheimer's disease |

# 4 Covariate definitions

Table 4: Covariate names, definitions and relevant codes for all variables included in main analyses

| Variable | Definition and codes |
| --- | --- |
| Age category | Age of patient at baseline |
| Education | Patient education prior to baseline; basic (elementary), medium (trade school; some college) or high (college +) |
| Income | Income range categorized into tertiles |
| Diabetes duration | Evidence of diabetes duration through diabetes diagnosis or evidence of metformin use; years from first evidence up until baseline |
| **Diagnoses** | Time-varying covariates; first evidence of ICD-08 or ICD-10 diagnosis code (below) prior to baseline or during follow up. Absence of diagnosis coded as 0 and presence as 1. |
| Heart failure | 42709, 42710, 42711, 42719, 42899, 78249, DI099, DI110, DI130, DI132, DI255, DI425, DI426, DI427, DI429, DI428A, DP290, DI43, DI50, DE105, DE115, DE125, DE135, DE145 |
| Renal disease | 403, 404, 580, 581, 582, 583, 584, 59009, 59319, 75310, 75311, 75312, 75313, 75314, 75315, 75316, 75317, 75318, 75319, 792, DN032, DN033, DN034, DN035, DN036, DN037, DN052, DN053, DN054, DN055, DN056, DN057, DZ490, DZ491, DZ492, DN18, DN19, DI120, DI131, DI132, DN250, DZ940, DZ992, DN26 |
| Chronic pulmonary disease | 515, 516, 517, 518, 490, 491, 492, 493, DJ40, DJ41, DJ42, DJ43, DJ44, DJ45, DJ46, DJ47, DJ60, DJ61, DJ62, DJ63, DJ64, DJ65, DJ66, DJ67, DJ684, DI278, DI279, DJ84, DJ701, DJ703, DJ920, DJ953, DJ961, DJ982, DJ983 |
| Any malignancy | 140, 141, 142, 143, 144, 145, 146, 147, 148, 149, 150, 151, 152, 153, 154, 155, 156, 157, 158, 159, 160, 161, 162, 163, 164, 165, 166, 167, 168, 169, 170, 171, 172, 174, 175, 176, 177, 178, 179, 180, 181, 182, 183, 184, 185, 186, 187, 188, 189, 190, 191, 192, 193, 194, 27559, DC0, DC1, DC2, DC3, DC40, DC41, DC42, DC43, DC44, DC45, DC46, DC47, DC48, DC49, DC5, DC6, DC70, DC71, DC72, DC73, DC74, DC75, DC76, DC86, DC97 |
| Ischemic heart disease | 411, 412, 413, 414, DI20, DI23, DI24, DI25 |
| Myocardial infarction | 410, DI21 |
| Hypertension | 41009, 41099, DI10, DI109, DI11, DI110, DI119, DI119A, DI12, DI120, DI129, DI129A, DI13, DI130, DI131, DI132, DI139, DI15, DI150, DI151, DI152, DI158, DI159, 401, 402, 403, 404 |
| Stroke | 433.434.435.436.437.438.DI63.DI64 |
| **Medications** | Time-varying covariates; in each time node, patients are classified as using the medication if a medication pickup with relevant ATC code (below) was made on a date falling in time bin. Absence of medication coded as 0 and presence as 1. |
| Beta-blockers | C07 |
| Calcium channel blockers | C08 |
| Renin-angiotensin system-acting inhibitors (RASI) | C09 |
| Loop diuretics | C03C, C03EB |
| Mineralocorticoid receptor antagonists (MRAs) | C03D |
| COPD medications | R03BA, R03AK06, R03AK07, R03AK08, R03AK09, R03AK10, R03AK11, R03AK12, R03AL08, R03AL09, R03AC, R03AK06, R03AK07, R03AK08, R03AK09, R03AK10, R03AK11, R03AK12, R03AK13, R03AL01, R03AL02, R03AL03, R03AL04, R03AL05, R03AL06, R03AL07, R03AL08, R03AL09, R03BB |

# 5 Simulation methods

We conducted a simulation study to find the best estimator for a longitudinal targeted maximum likelihood estimation (L-TMLE) analysis of the effect of second-line diabetes drugs on dementia risk. The data was simulated to approximate the features and multivariate distribution of the Danish National Registry analysis dataset used for the primary analysis, which has five key features: many time points (10), a relatively rare treatment regime of interest, a rare outcome and exposure , competing risks from death, and a high degree of administrative censoring. Two simulation scenarios were used:

1. A simulation using data with a similar multivariate distribution of treatment regimes $A$, and covariates $L$, and outcomes $Y$ to the Danish registry data. The truth was calculated as the counterfactual 5-year risk of dementia prior to death when continuously on GLP-1RA's versus not, with the effect of GLP-1RA's on death removed to eliminate the competing risk.

2. A simulation using data with a similar distribution of treatment regimes $A$ and covariates $L$ to the Danish registry data but with permuted outcomes $Y$ such that there was no association between $A$ and $Y$ and the true treatment effect is null.

For both simulation scenarios, we estimated the effect of 5-year continuous usage of GLP-1RA compared to no GLP-1RA usage, analogous to the target causal parameter 1 (Table 1 of main text, column 1).

Fore each of the two data generating processes outlined above, we examined:

- *Estimator selection*: We considered both inverse-probability of treatment weighting (IPTW) and targeted maximum likelihood estimation (TMLE)

- *Nuisance parameter estimation*:
  We considered in the simulation the following candidate algorithms for g and Q estimation:
    - LASSO ($\lambda$ (lambda) chosen to minimize cross-validated error, and undersmoothed*)
    - Ridge ($\lambda$ chosen to minimize cross-validated error, and undersmoothed*)
    - Elastic net ($\alpha = 0.5$ and $\lambda$ chosen to minimize cross-validated error, and undersmoothed*)
    - Generalized linear models (identity link)
    - Random forest

- For each of the penalized regressions above (LASSO, Ridge, Elastic net) we considered one with a cross-validated tuning parameter selection, and one undersmoothed version which utilized the minimum $\lambda$ value (i.e. the least penalization) that did not fail across a range of candidate values.

- *Variance estimation*: To estimate variance, we considered empirical influence curve variance, "robust" variance estimator [2], and the non-parametric bootstrap.

We estimated the performance of each combination of the above specifications on 500 datasets generated from our data-generating process described above. To evaluate performance of the estimation procedures, we compared oracle coverage, estimator bias ($\sigma^2$), and estimator variance of the risk difference (see Section 7). These quantities were calculated using the true parameter value ($\Psi_0$), the estimated parameter value ($\hat{\psi}_i$) for each iteration of the simulated data $i$ where $i = 1....500$:

Bias: $E(\hat{\psi}) - \Psi_0$

Variance: $E[\hat{\psi}^2] - [E(\hat{\psi})]^2$

Bias/SE ratio: $(E(\hat{\psi}) - \Psi_0)/(E[\hat{\psi}^2] - [E(\hat{\psi})])$

Oracle Coverage: $E(I_{OCI_{lb} \leq \Psi_0 \leq OCI_{ub}})$

Where the Oracle upper ($OCI_{ub}$) and lower bound ($OCI_{lb}$) are calculated using the standard deviation of the estimator across simulation repetitions: $\Psi_i \pm 1.96\sigma_{est}$

Code used for the simulation is available on Github.[1]

# 6 Simulation results

## 6.1 Scenario 1: Realistic simulation, protective effect of GLP1 on dementia

True disk difference -0.0.00719; Outcome prevalence 0.83% in treated group, 1.55% in control group

### 6.1.1 Performance of difference point estimators (scenario 1)

| Estimator | Algorithm | Lambda | Truncation | RD bias | RD variance | RD bias SE ratio | RD oracle 95% coverage |
|---|---|---|---|---|---|---|---|
| **TMLE** | **Ridge** | **Undersmoothed** | **Less than 0.01** | **0.001851** | **5.0e-06** | **0.835818** | **95.00000** |
| TMLE | Elastic Net | Undersmoothed | Less than 0.01 | 0.001963 | 4.0e-06 | 0.981080 | 95.60000 |
| TMLE | LASSO | Undersmoothed | Less than 0.01 | 0.001964 | 4.0e-06 | 0.980587 | 95.60000 |
| TMLE | Ridge | Undersmoothed | Untruncated | 0.002251 | 1.1e-05 | 0.691243 | 96.40000 |
| TMLE | Elastic Net | Undersmoothed | Untruncated | 0.002478 | 9.0e-06 | 0.829886 | 96.20000 |
| TMLE | LASSO | Undersmoothed | Untruncated | 0.002484 | 9.0e-06 | 0.827790 | 96.20000 |
| TMLE | Elastic Net | CV-minimum SE | Untruncated | 0.002538 | 3.0e-06 | 1.549389 | 73.40000 |
| TMLE | Elastic Net | CV-minimum SE | Less than 0.01 | 0.002544 | 3.0e-06 | 1.585283 | 71.00000 |
| TMLE | Ridge | CV-minimum SE | Untruncated | 0.002551 | 4.0e-06 | 1.207114 | 76.20000 |
| TMLE | LASSO | CV-minimum SE | Untruncated | 0.002558 | 3.0e-06 | 1.585686 | 73.00000 |
| TMLE | LASSO | CV-minimum SE | Less than 0.01 | 0.002565 | 3.0e-06 | 1.618536 | 70.20000 |
| TMLE | Ridge | CV-minimum SE | Less than 0.01 | 0.002614 | 4.0e-06 | 1.356658 | 68.60000 |
| TMLE | Random Forest | N/A | Less than 0.01 | 0.002771 | 1.0e-06 | 2.325057 | 92.00000 |
| IPTW | Ridge | Undersmoothed | Less than 0.01 | 0.004216 | 5.0e-06 | 1.870010 | 54.80000 |
| IPTW | LASSO | Undersmoothed | Less than 0.01 | 0.004216 | 4.0e-06 | 2.070313 | 54.80000 |
| IPTW | Elastic Net | Undersmoothed | Less than 0.01 | 0.004216 | 4.0e-06 | 2.072648 | 54.80000 |
| IPTW | Ridge | CV-minimum SE | Less than 0.01 | 0.004216 | 4.0e-06 | 2.166498 | 54.80000 |
| IPTW | Elastic Net | CV-minimum SE | Untruncated | 0.004216 | 3.0e-06 | 2.556702 | 54.80000 |
| IPTW | LASSO | CV-minimum SE | Untruncated | 0.004216 | 3.0e-06 | 2.596233 | 54.80000 |
| IPTW | Elastic Net | CV-minimum SE | Less than 0.01 | 0.004216 | 3.0e-06 | 2.613575 | 54.80000 |
| IPTW | LASSO | CV-minimum SE | Less than 0.01 | 0.004216 | 3.0e-06 | 2.647554 | 54.80000 |
| IPTW | Ridge | CV-minimum SE | Untruncated | 0.004327 | 5.0e-06 | 2.026333 | 53.77778 |
| IPTW | Ridge | Undersmoothed | Untruncated | 0.004331 | 1.1e-05 | 1.290068 | 53.58491 |
| IPTW | Random Forest | N/A | Less than 0.01 | 0.004346 | 1.0e-06 | 3.864224 | 53.60000 |
| IPTW | LASSO | Undersmoothed | Untruncated | 0.004468 | 9.0e-06 | 1.453555 | 50.60606 |
| IPTW | Elastic Net | Undersmoothed | Untruncated | 0.004468 | 9.0e-06 | 1.462082 | 50.60606 |

The estimator chosen based on oracle coverage closest to 95% with the lowest bias and variance is bolded and highlighted with lines. Here, for the risk difference with data scenario 1, the method chosen from the simulation results is L-TMLE using undersmoothed ridge regression with g-truncation. Importantly, the undersmoothed option for various algorithms tend to perform better than the default lambda parameter selection.

### 6.1.2 Performance of difference variance estimators (scenario 1)

| Variance.estimator | RD Coverage |
|---|---|
| Influence curve | 89.2 |
| Bootstrap | 94.8 |
| TMLE | 99.8 |

Comparing different variance estimators using the estimator and estimation method chosen above, we find that the influence curve (ic) method is anti-conservative, and the TMLE method is conservative, with very wide confidence interval widths. The bootstrap estimator performs the best of the different variance options.

## 6.2 Scenario 2: Realistic simulation, no exposure-outcome relationship

The second simulation was done to benchmark different estimator performance using data with a realistic prevalence of dementia and 5 years of follow-up, but with a known null association (True Risk Difference=0) between treatments and the dementia outcome.

### 6.2.1 Comparison of different estimators' performance (scenario 2)

| Estimator | Algorithm | Lambda | Truncation | RD bias | RD variance | RD bias SE ratio | RD oracle 95% coverage |
|---|---|---|---|---|---|---|---|
| TMLE | LASSO | CV-minimum SE | Less than 0.01 | 0.001599 | 3e-06 | 0.901633 | 96.0 |
| TMLE | LASSO | CV-minimum SE | Untruncated | 0.001599 | 3e-06 | 0.901413 | 96.0 |
| TMLE | Elastic Net | CV-minimum SE | Less than 0.01 | 0.001602 | 3e-06 | 0.884646 | 96.0 |
| TMLE | Elastic Net | CV-minimum SE | Untruncated | 0.001602 | 3e-06 | 0.884224 | 96.0 |
| TMLE | Ridge | CV-minimum SE | Less than 0.01 | 0.001615 | 6e-06 | 0.663841 | 95.4 |
| TMLE | Ridge | CV-minimum SE | Untruncated | 0.001629 | 6e-06 | 0.662149 | 95.8 |
| **TMLE** | **Ridge** | **Undersmoothed** | **Less than 0.01** | **0.001689** | **6e-06** | **0.688838** | **94.4** |
| TMLE | Elastic Net | Undersmoothed | Less than 0.01 | 0.001736 | 4e-06 | 0.912383 | 94.4 |
| TMLE | LASSO | Undersmoothed | Less than 0.01 | 0.001737 | 4e-06 | 0.913133 | 94.4 |
| TMLE | Ridge | Undersmoothed | Untruncated | 0.001738 | 7e-06 | 0.677415 | 95.0 |
| TMLE | Elastic Net | Undersmoothed | Untruncated | 0.001819 | 4e-06 | 0.886583 | 94.8 |
| TMLE | LASSO | Undersmoothed | Untruncated | 0.001821 | 4e-06 | 0.886609 | 94.8 |
| IPTW | Ridge | Undersmoothed | Untruncated | 0.002642 | 7e-06 | 1.014801 | 88.2 |
| IPTW | Ridge | CV-minimum SE | Untruncated | 0.002642 | 6e-06 | 1.058681 | 88.2 |
| IPTW | Ridge | Undersmoothed | Less than 0.01 | 0.002642 | 6e-06 | 1.064925 | 88.2 |
| IPTW | Ridge | CV-minimum SE | Less than 0.01 | 0.002642 | 6e-06 | 1.071845 | 88.2 |
| IPTW | LASSO | Undersmoothed | Untruncated | 0.002642 | 4e-06 | 1.266778 | 88.2 |
| IPTW | Elastic Net | Undersmoothed | Untruncated | 0.002642 | 4e-06 | 1.268788 | 88.2 |
| IPTW | Elastic Net | Undersmoothed | Less than 0.01 | 0.002642 | 4e-06 | 1.376521 | 88.2 |
| IPTW | LASSO | Undersmoothed | Less than 0.01 | 0.002642 | 4e-06 | 1.377126 | 88.2 |
| IPTW | Elastic Net | CV-minimum SE | Untruncated | 0.002642 | 3e-06 | 1.443315 | 88.2 |
| IPTW | Elastic Net | CV-minimum SE | Less than 0.01 | 0.002642 | 3e-06 | 1.444487 | 88.2 |
| IPTW | LASSO | CV-minimum SE | Untruncated | 0.002642 | 3e-06 | 1.474387 | 88.2 |
| IPTW | LASSO | CV-minimum SE | Less than 0.01 | 0.002642 | 3e-06 | 1.475018 | 88.2 |

In Scenario 2, g truncation does not make a sizeable difference for CV-min SE estimators with penalized regression. The undersmoothed truncated penalized regression algorithms here perform similarly.

# 7 Baseline demographic characteristics by baseline treatment

**SGLT2i**

| Variable | Level | SGLT2i (n=22434) | no SGLT2i (n=82494) | Total (n=104928) |
|---|---|---|---|---|
| Age groups | Below 55 | 3,329 (14.8) | 11,628 (14.1) | 14,957 (14.3) |
| | 55-60 | 4013 (17.9) | 13260 (16.1) | 17,273 (16.5) |
| | 60-65 | 4164 (18.6) | 14361 (17.4) | 18,525 (17.7) |
| | 65-70 | 3769 (16.8) | 14231 (17.3) | 18,000 (17.2) |
| | 70-75 | 3526 (15.7) | 12109 (14.7) | 15,635 (14.9) |
| | 75-80 | 2296 (10.2) | 8505 (10.3) | 10,801 (10.3) |
| | 80-85 | 1003 (4.5) | 5126 (6.2) | 6,129 (5.8) |
| | 85-90 | 269 (1.2) | 2351 (2.8) | 2,620 (2.5) |
| | Above 90 | 65 (0.3) | 923 (1.1) | 988 (0.9) |
| Education | Basic | 8,022 (35.8) | 35,376 (42.9) | 43,398 (41.4) |
| | Medium | 11113 (49.5) | 36401 (44.1) | 47,514 (45.3) |
| | High | 3299 (14.7) | 10717 (13.0) | 14,016 (13.4) |
| Income (tertile) | Q1 | 5,757 (25.7) | 29,269 (35.5) | 35,026 (33.4) |
| | Q2 | 7163 (31.9) | 27788 (33.7) | 34,951 (33.3) |
| | Q3 | 9514 (42.4) | 25437 (30.8) | 34,951 (33.3) |
| Diabetes duration (years) | Below 1 | 5,489 (24.5) | 30,053 (36.4) | 35,542 (33.9) |
| | 1-5 | 8193 (36.5) | 30473 (36.9) | 38,666 (36.9) |
| | 5-10 | 6501 (29.0) | 17513 (21.2) | 24,014 (22.9) |
| | Above 10 | 2251 (10.0) | 4455 (5.4) | 6,706 (6.4) |
| Diagnoses | | | | |
|   Heart failure | Yes | 1599 (7.1) | 4901 (5.9) | 6,500 (6.2) |
|   Renal disease | Yes | 138 (0.6) | 1087 (1.3) | 1,225 (1.2) |
|   Chronic pulmonary disease | Yes | 1224 (5.5) | 5477 (6.6) | 6,701 (6.4) |
|   Any malignancy | Yes | 1316 (5.9) | 5739 (7.0) | 7,055 (6.7) |
|   Ischemic heart disease | Yes | 2635 (11.7) | 10091 (12.2) | 12,726 (12.1) |
|   Myocardial infarction | Yes | 1561 (7.0) | 4263 (5.2) | 5,824 (5.6) |
|   Hypertension | Yes | 4573 (20.4) | 18552 (22.5) | 23,125 (22.0) |
|   Stroke | Yes | 929 (4.1) | 3651 (4.4) | 4,580 (4.4) |
| Medications | | | | |
|   Beta blockers | Yes | 7047 (31.4) | 23793 (28.8) | 30,840 (29.4) |
|   Calcium channel blockers | Yes | 6938 (30.9) | 24274 (29.4) | 31,212 (29.7) |
|   Renin-angiotensin system-acting inhibitors | Yes | 14811 (66.0) | 51451 (62.4) | 66,262 (63.1) |
|   Loop diuretics | Yes | 3114 (13.9) | 13536 (16.4) | 16,650 (15.9) |
|   Mineralocorticoid receptor antagonists | Yes | 1797 (8.0) | 4595 (5.6) | 6,392 (6.1) |
|   COPD medications | Yes | 3002 (13.4) | 10887 (13.2) | 13,889 (13.2) |

**DPP4**

| Variable | Level | DPP4 (n=40569) | no DPP4 (n=64359) | Total (n=104928) |
|---|---|---|---|---|
| Age groups | Below 55 | 5,438 (13.4) | 9,519 (14.8) | 14,957 (14.3) |
| | 55-60 | 6326 (15.6) | 10947 (17.0) | 17,273 (16.5) |
| | 60-65 | 7009 (17.3) | 11516 (17.9) | 18,525 (17.7) |
| | 65-70 | 6998 (17.2) | 11002 (17.1) | 18,000 (17.2) |
| | 70-75 | 6145 (15.1) | 9490 (14.7) | 15,635 (14.9) |
| | 75-80 | 4327 (10.7) | 6474 (10.1) | 10,801 (10.3) |
| | 80-85 | 2633 (6.5) | 3496 (5.4) | 6,129 (5.8) |
| | 85-90 | 1224 (3.0) | 1396 (2.2) | 2,620 (2.5) |
| | Above 90 | 469 (1.2) | 519 (0.8) | 988 (0.9) |
| Education | Basic | 16,919 (41.7) | 26,479 (41.1) | 43,398 (41.4) |
| | Medium | 18331 (45.2) | 29183 (45.3) | 47,514 (45.3) |
| | High | 5319 (13.1) | 8697 (13.5) | 14,016 (13.4) |
| Income (tertile) | Q1 | 13,884 (34.2) | 21,142 (32.9) | 35,026 (33.4) |
| | Q2 | 14095 (34.7) | 20856 (32.4) | 34,951 (33.3) |
| | Q3 | 12590 (31.0) | 22361 (34.7) | 34,951 (33.3) |
| Diabetes duration (years) | Below 1 | 11,377 (28.0) | 24,165 (37.5) | 35,542 (33.9) |
| | 1-5 | 16287 (40.1) | 22379 (34.8) | 38,666 (36.9) |
| | 5-10 | 10307 (25.4) | 13707 (21.3) | 24,014 (22.9) |
| | Above 10 | 2598 (6.4) | 4108 (6.4) | 6,706 (6.4) |
| Diagnoses | | | | |
| Heart failure | Yes | 2445 (6.0) | 4055 (6.3) | 6,500 (6.2) |
| Renal disease | Yes | 626 (1.5) | 599 (0.9) | 1,225 (1.2) |
| Chronic pulmonary disease | Yes | 2661 (6.6) | 4040 (6.3) | 6,701 (6.4) |
| Any malignancy | Yes | 2953 (7.3) | 4102 (6.4) | 7,055 (6.7) |
| Ischemic heart disease | Yes | 4799 (11.8) | 7927 (12.3) | 12,726 (12.1) |
| Myocardial infarction | Yes | 1991 (4.9) | 3833 (6.0) | 5,824 (5.6) |
| Hypertension | Yes | 9215 (22.7) | 13910 (21.6) | 23,125 (22.0) |
| Stroke | Yes | 1788 (4.4) | 2792 (4.3) | 4,580 (4.4) |
| Medications | | | | |
| Beta blockers | Yes | 11801 (29.1) | 19039 (29.6) | 30,840 (29.4) |
| Calcium channel blockers | Yes | 11979 (29.5) | 19233 (29.9) | 31,212 (29.7) |
| Renin-angiotensin system-acting inhibitors | Yes | 25598 (63.1) | 40664 (63.2) | 66,262 (63.1) |
| Loop diuretics | Yes | 6553 (16.2) | 10097 (15.7) | 16,650 (15.9) |
| Mineralocorticoid receptor antagonists | Yes | 2162 (5.3) | 4230 (6.6) | 6,392 (6.1) |
| COPD medications | Yes | 5319 (13.1) | 8570 (13.3) | 13,889 (13.2) |

# 8 Supplemental analysis: time discretization

| Analysis | Treatment | Control | RD% (rescaled x100, 95 % CI) | RR (95 % CI) |
|---|---|---|---|---|
| 6-month intervals (main analysis) | 0.0093 | 0.0141 | -0.47(-0.94,-0.01 ) | 0.66 ( 0.40, 0.96) |
| 3-month intervals | 0.009 | 0.014 | -0.50 (-1.01, -0.002) | 0.63 (0.37, 1.12) |
| 12-month intervals | 0.010 | 0.015 | -0.47 (-0.91,-0.03) | 0.68 (0.43, 1.06) |

Table 8: Five-year risk difference of dementia comparing sustained GLP1 use vs. sustained non-use in a cohort of Danish diabetes patients initiating second-line therapy under distinct time discretization

# 9 Relative risk results

| Comparison | Treatment Risk | Control Risk | Risk Ratio (RR, 95% CI) |
|---|---|---|---|
| 1) GLP1 vs. no GLP1 | 0.0093 | 0.0141 | 0.66 ( 0.40, 0.96) |
| 2) GLP1 vs. SGLT2 | 0.0071 | 0.0074 | 0.97 (0.43, 2.00) |
| 3) GLP1 vs. DPP4 | 0.0068 | 0.0115 | 0.74 (0.42, 1.04) |

Table 9: Five-year risk ratio of dementia under sustained GLP1-RA use vs sustained comparator use in a cohort of Danish diabetes patients initiating second-line therapy


\end{document}